%% file: main.tex
\documentclass[manuscript]{acmart}

\AtBeginDocument{%
  }

\acmYear{2024}\copyrightyear{2024}
\acmConference[ICTD '24]{International Conference on Information \& Communication Technologies and Development}{December 9--11, 2024}{Nairobi, Kenya}
\acmBooktitle{International Conference on Information \& Communication Technologies and Development (ICTD '24), December 9--11, 2024, Nairobi, Kenya}
\acmDOI{10.1145/3700794.3700818}
\acmISBN{979-8-4007-1041-4/24/12}

\usepackage{xspace}
\usepackage{subcaption}
\usepackage{url}
\usepackage{hyperref}
\newcommand{\name}{{GAIUS}\xspace}

\newcommand*\blue{\color{black}}

\begin{document}

\title{The \name Experience: Powering a Hyperlocal Mobile Web for Communities in Emerging Regions}

% \affiliation{%

\author{Rohail Asim}
\email{rohail.asim@nyu.edu}
\affiliation{
    \institution{New York University Abu Dhabi}
    \country{UAE}
}

\author{Arjuna Sathiaseelan}
\affiliation{%
  \institution{Cambridge}
  \country{UK}
}

\author{Arko Chatterjee}
\affiliation{%
 \institution{Gaius Networks}
 \country{USA}
 }
 
\author{Mukund Lal}
\affiliation{%
  \institution{Gaius Networks}
  \country{USA}
}

\author{Yasir Zaki}
% \authornote{Note}
\affiliation{%
  \institution{New York University Abu Dhabi}
    \country{UAE}
}
\email{yasir.zaki@nyu.edu}

\author{Lakshmi Subramanian}
% \authornote{Note}
\affiliation{%
  \institution{New York University}
    \country{USA}
}
\email{lakshmi@nyu.edu}

\renewcommand{\shortauthors}{Asim et al.}

\begin{abstract}
Despite increasing mobile Internet penetration in developing regions, mobile users continue to experience a poor web experience due to two key factors: (i) lack of locally relevant content; (ii) poor web performance due to complex web pages and poor network conditions. In this paper, we describe our design, implementation and deployment experiences of \name, a mobile content ecosystem that enables efficient creation and dissemination of locally relevant web content into hyperlocal communities in emerging markets. The basic building blocks of \name are a lightweight content edge platform combined with a mobile application that collectively provide a Hyperlocal Web abstraction for mobile users to create and consume locally relevant content and interact with other users via a community abstraction.  The \name platform uses MAML, a web specification language that dramatically simplifies web pages to reduce the complexity of Web content within the \name ecosystem, improve page load times and reduce network costs. In this paper, we describe our experiences deploying \name across a large user base in India, Bangladesh and Kenya.
\end{abstract}





\maketitle

\input{intro}

\input{relatedwork}
\input{gaius}
\input{implementation}

\input{eval}
\input{deployment}
\input{conclusion}
\bibliographystyle{ACM-Reference-Format}
\bibliography{references}


\end{document}

%% file: intro.tex
\section{Introduction}

Many challenges constrain Web access in developing regions. A classic performance related challenge is the unnecessary complexity of web pages where a browser typically: (a) downloads $100+$ objects~\cite{webcomplexity} per page; (b) spawns $30+$ network connections~\cite{Sundaresan:2013:CCA:2504730.2504741, elkhatib2014can}; (c) issues $20+$ DNS requests~\cite{Sundaresan:2013:CCA:2504730.2504741}; and (d) processes several layers of recursive requests triggered by JavaScript and HTTP re-directions~\cite{webcomplexity}. Together, these factors collectively lead to poor web performance in constrained network conditions in emerging markets~\cite{Chen:2011:AAW:1963192.1963358,zaki2014dissecting}. Performance factors aside, one major problem that has remained less explored is the lack of cheap, locally relevant content. This shortage of relevant local content makes the Internet less attractive for users in developing regions~\cite{ICANN}\cite{KPMG}.

To address these problems, this paper describes our experiences in the design, implementation and deployment of \name, a hyperlocal mobile web ecosystem for local content creation and diffusion at the edge. {\blue This means that the ecosystem is specialized to enable small localized communities to efficiently create and consume locally relevant content, and the system infrastructure to provide these services is distributed across many nodes in order to minimize the distances between any localized community and its nearest node}. The design of \name addresses five key pain-points that has limited the reach and relevance of the Web in emerging regions. First, \name provides a highly simplified mobile interface that enables users with limited technical expertise to easily create, consume and disseminate rich content with local language support. Second, \name provides a hyperlocal web abstraction that enables users in a locality to easily interact and transact within themselves via three core elements: content, communities (public and private) and a marketplace. Third, the \name architecture is decentralized {\blue i.e. the workload for providing the \name ecosystem's services to its end users is handled by multiple nodes that are in close proximity to the end user, as opposed to one central server with limited scalability and fault tolerance}. This local edge deployment model empowers any local operator to create and sustain a local \name ecosystem  for a local community in addition to reducing end-to-end latencies for page loads. Fourth, to simplify the complexity of pages, \name pages are represented in a highly concise Mobile Application Markup Language (MAML) format instead of HTML and JavaScript that enables \name to operate efficiently in bandwidth constrained environments. 
Finally, to make the hyperlocal Web ecosystem sustainable, \name provides a local content-ad exchange that empowers local and global advertisers to advertise within a hyperlocal ecosystem.

Architecturally, \name consists of a set of locally distributed servers at the edge that host \name pages~(interchangeably referred to as channels) and an application running on user's smartphone device that enables consumption and creation of content. When a user requests a page, a \name policy-engine takes into account user's preferences, content provider's policies, and advertiser's contracts to deliver the correct page and advertisements. The \name pages are represented in a highly concise Mobile Application Markup Language (MAML) format instead of HTML and JavaScript, to minimize communication overhead. Finally, the \name app on an end user's device receives, interprets, and renders the MAML pages. The \name application is also used to create content and publish it on the \name servers. Together, these components enable the easy creation, diffusion, and consumption of content. To enable payments for services, \name has an integrated micro-payment scheme, which is already a prevalent payment model in many developing countries~\cite{wishart2006micro}. 

% In summary, the \name ecosystem is designed around a number of key concepts:

% \begin{itemize}
% 	\item Hosting and diffusing content close to the users.
% 	\item Coordinating a mobile application and an edge server to efficiently view and create content.
% 	\item Rewriting pages using an efficient Mobile Application Markup Language (MAML).
% 	\item Enabling a content and ad exchange marketplace.
% \end{itemize}

This paper describes our experiences deploying \name to support a hyperlocal mobile web abstraction for a large user base spread across several communities in Bangladesh, India and Kenya.  Our deployments at scale with 100K users demonstrate that \name enables users (even with limited Web experience) to easily create and publish local content, interact seamlessly within communities, advertise their products and tailor their hyperlocal experience as a function of their interests and network conditions. We describe our experiences powering educational communities in India, entertainment-centric communities in Bangladesh and small business communities in Kenya. From a performance perspective, we show that a \name page loads faster than the regular page (up to 60\%), while reducing the overall number of objects requested (by 80\%), and consequently the overall page size (up to 57\%). %In particular, we show that the median page load time reduction is 80\% and the median page size reduction is 60\%. \name is also able to reduce the number of web requests by 80\%. Additionally, in comparison to other related state-of-the-art solutions such as OperaMini, we demonstrate that the \name ecosystem is capable of delivering pages 20x faster, and is able to reduce the size of pages by up to 57\%.

%% file: relatedwork.tex
\section{Related Work}
We start by highlighting the key factors behind the poor performance for bandwidth-constrained users. We discuss various academic and commercial solutions that are proposed to fix these problems. We also outline how current content delivery infrastructure is non-conducive to dissemination of localized content.

\noindent\textbf{Web Performance and Complexity:}
To overcome the challenges of poor web performance in developing countries, several systems and techniques have been proposed in recent years to optimize web browsing over poor network connections including network-level optimizations, caching techniques and content distribution mechanisms~\cite{Isaacman_thec-link, Thies02searchingthe, Chetty:2011:WMI:1978942.1979217, Chen:2009:RWS:1526709.1526765, Chen:2011:DIC:1963192.1963359, Chen:2011:AAW:1963192.1963358,xcache10.1145/3136560.3136577}. Recent works~\cite{klotski,shandian,netravali2016polaris} have focused on the complexity of web pages and suggested different approaches to address this problem. Klotski~\cite{klotski}, for example, proposed re-prioritizing web content relevant to a user's preferences by enabling fast selection and load-time estimation for the subset of resources to be prioritized. Another recent system, Shandian~\cite{shandian}, uses a split-browser architecture to restructure the web page load process and control the portions of the web page that get communicated. They chose a split-browser architecture to deploy the solution. {\blue xCache~\cite{xcache10.1145/3136560.3136577} is a cloud-managed Internet caching architecture that aims to maintain the liveness of popular content at software defined edge caches in developing regions and is able to work in conjunction with optimization solutions like Shandian}. Polaris~\cite{netravali2016polaris} uses fine-grained object dependency graphs to quickly load objects browser rendering. {\blue Finally, Chaqfeh et al.\ showed that developing regions rely heavily on the usage of low-end devices that, in addition to the poor network conditions, further exacerbate poor web performance in these regions~\cite{digital_divide}.}

\noindent\textbf{Content Rewriting and Simplification:} 
There have been many efforts proposing new methodologies to rewrite web pages for mobile users that support heterogeneous devices~\cite{Kulkarni:2011:AAW:1979742.1979810} and bandwidth-constrained users~\cite{Chen:2011:DIC:1963192.1963359,Chen:2009:RWS:1526709.1526765}. {\blue Slimweb~\cite{chaqfehslimweb10.1145/3572334.3572397} creates lightweight versions of mobile web pages on-the-fly by eliminating non-essential JavaScript. Also solutions like JSCleaner~\cite{JSCleaner}, JSAnalyzer~\cite{JSanalyzer}, and Muzeel~\cite{Muzeel} focused on improving the page load time and simplifying the page complexity by either removing unnecessary JavaScript files or de-cluttering their unused code.}
Kumar et al.~\cite{Kumar:2009:ARW:1520340.1520646, kumar2012data, Kumar:2013:WDM:2470654.2466420, Kumar:2011:BER:1978942.1979262} have proposed different techniques to understand the DOM structure of web pages and re-design a web page to different layouts. This line of work produces pages that can be tailored for different domains including mobile users. For e.g. data-driven web design~\cite{kumar2012data} uses machine learning applications for rapid re-targeting between page designs. Using this application, a web page designed for desktop can be automatically re-targeted for mobile devices. To enable support for such data driven web design tools, a design mining platform called Webzeitgeist~\cite{Kumar:2013:WDM:2470654.2466420} supports a repository of over 100,000 web pages and 100 million design elements. Bricolage~\cite{Kumar:2011:BER:1978942.1979262} learns how to transfer content between pages with different design templates for rapid re-targeting between page designs. Within ICTD contexts, the most closely related work is RuralCafe~\cite{Chen:2009:RWS:1526709.1526765}, which supports different fidelity levels for low-bandwidth or high-latency connections through filtering of content types and a split-browser architecture.

\noindent\textbf{Existing Commercial Solutions:} 
In the context of developing regions, recent commercial attempts have been made by Google AMP~\cite{googleAMP}, Facebook Instant articles~\cite{facebook} and Opera mobile browser~\cite{opera}. AMP introduced new HTML tags and elements that are used to manage resource loading. AMP enables parallel download of objects by leveraging iframes and by declaring the size/positions of objects upfront. AMP also uses caches to ensure that pages satisfy all AMP specifications and enable pre-rendering of pages. 
% AMP caches provide the point where other optimizations such as reordering HTML tags, rewriting JavaScript URLs, compression and other image operations are performed. 
Facebook Instant Articles is a platform within mobile Facebook application, created by Facebook for publishers to create fast and interactive articles on Facebook. Facebook claims instant articles make the loading of articles ten times faster than loading of standard mobile articles. Facebook enables faster access of instant articles by preloading the data. 
% On tapping the article, only very small data that is not previously pre-loaded is downloaded and hence the article loads faster than a standard mobile article. 
Opera Mobile browser and its lower-end sibling, Opera Mini browser, request web pages through a proxy server called Opera Turbo. Opera Turbo compresses web pages by about 90\% rendering them faster by two to three times. Opera has another extreme compression method called Mini mode. It has higher compression ratio than Turbo mode, but with a loss of certain functionality of web page rendered. 
% Also Opera mini browser has in-built adBlocker which further prunes content sent to the mobile phone. As per August 2017, most of the market for Opera browsers is from developing countries with slower internet like India, Bangladesh, Indonesia, South Africa, Nigeria, Ukraine, Hungary, Kuwait, Tanzania and Pakistan.
Neighbourly~\cite{neighbourly} and  Nextdoor~\cite{nextdoor} are commercial solutions that aim at helping users engage directly with their neighbourhoods. Facebook communities~\cite{FBcomm}, WhatsApp~\cite{whatsapp}, WeChat~\cite{wechat}, GAIUS~\cite{gaiusmobicom, gaiusictd} and ShareChat~\cite{sharechat} are other examples of commercial solutions that revolve around engaging mobile users in communities and groups.

{\blue CGNet Swara~\cite{mudliar_thies10.1145/2160673.2160695} is another local ecosystem that aims to amplify the voices of rural communities and overcome challenges induced by the under-representation of these communities in the mainstream media. The TEK search engine~\cite{levison2001the} goes a step further and attempts to enable users to access Web resources and information using only email, to support rural communities where internet access and bandwidth is extremely restrictive. Awaaz~\cite{singhawaaz10.1145/3136560.3136568} is a citizen journalism portal that leverages Fracebook's Free Basics program to deploy a service that allows users in developing regions to report local issues with the location and images.}

The design philosophy of \name fundamentally differs from many of these commercial solutions in that it promotes a decentralized hyperlocal mobile Web ecosystem with a different mobile app and edge abstraction with support for rich local content, hyperlocal community interactions, marketplace interactions, local advertising, edge deployment model and local ownership. \name also has a fundamentally different data ownership model that enable local ecosystem providers to own and manage their own local \name deployment instances. \name also primarily focuses on mobile users in emerging markets which motivates the use of the MAML web representation format for enhancing performance in constrained networks. This experience paper significantly expands upon the hyperlocal Web vision and describes our commercial deployment experiences in rolling out the \name system in three emerging markets: Bangladesh, India and Kenya.

%in that, \name aims to help users in developing regions to create their own ecosystem of content and advertisements. First, from the end-user perspective \name allows users to simplify the expression and discoverability of new content. Second, unlike previous works, \name explicitly includes advertisers within its ecosystem. Third, like prior work, \name, improves web performance of `remote' web content, but does so through a simplified DOM representation that explicitly eliminates over-reliance of HTML/Javascript/CSS on recursive handling of objects while maintaining functional equivalence with the original page. None of the previously mentioned solutions tackle these fundamental issues. 
% While Google AMP comes really close to simplifying the web page and doing several optimizations, it still relies on HTML, JS and CSS heavily. In addition to reducing the dependence on HTML/JavaScript \name also proposes a new way to deliver the web pages of correct fidelity to the user based on their bandwidth measurement. 

\noindent\textbf{Content Delivery Networks:}
Content distribution networks (CDN), which traditionally aim to replicate relevant content in the network edges~\cite{akamai,cloudflare} are ineffective for enhancing performance in the developing world~\cite{Sharma:2015:RSC:2830629.2830649,Bischof:2015:OCC:2815675.2815718,fanou2016pushing}. The underlying factors are three-fold: poor connectivity~\cite{feamster}, high latencies, and web complexity. Even with heavily condensed web pages (without JavaScript, high resolution images, videos or iFrames) as offered by Facebook's Free Basics content delivery performance is poor~\cite{sen2016free}. %In addition, CDNs do not provide any mechanisms for local content and advertisement providers, but it has been shown that native content shapes how users experience content~\cite{wojdynski2016going}. CDNs are also predominantly designed to be agnostic to the end-user device~\cite{ahmad2016view}.
% , but low-cost phones are used extensively in developing regions. 

%% file: gaius.tex
\section{The \name Ecosystem}
The \name ecosystem focuses on engaging users via hyper-local communities powered by hyper-local content, ads, promotions and services (e.g local marketplace) for each community thereby enabling interactions and transactions among users within communities in their own languages. %\name provides a unified platform that enables a creation of broad spectra of hyper-local communities across different languages for urban/rural users in emerging markets. %\name has identified three key types of content-driven communities:
%\begin{enumerate}
%	\item educational communities in colleges and schools with jobs, tutor services and product offerings
%	\item local businesses offering marketplace for residents
%	\item local news, media, entertainment for communities
%\end{enumerate}
\begin{figure}[hbt]
\centering
  \includegraphics[width=0.7\columnwidth]{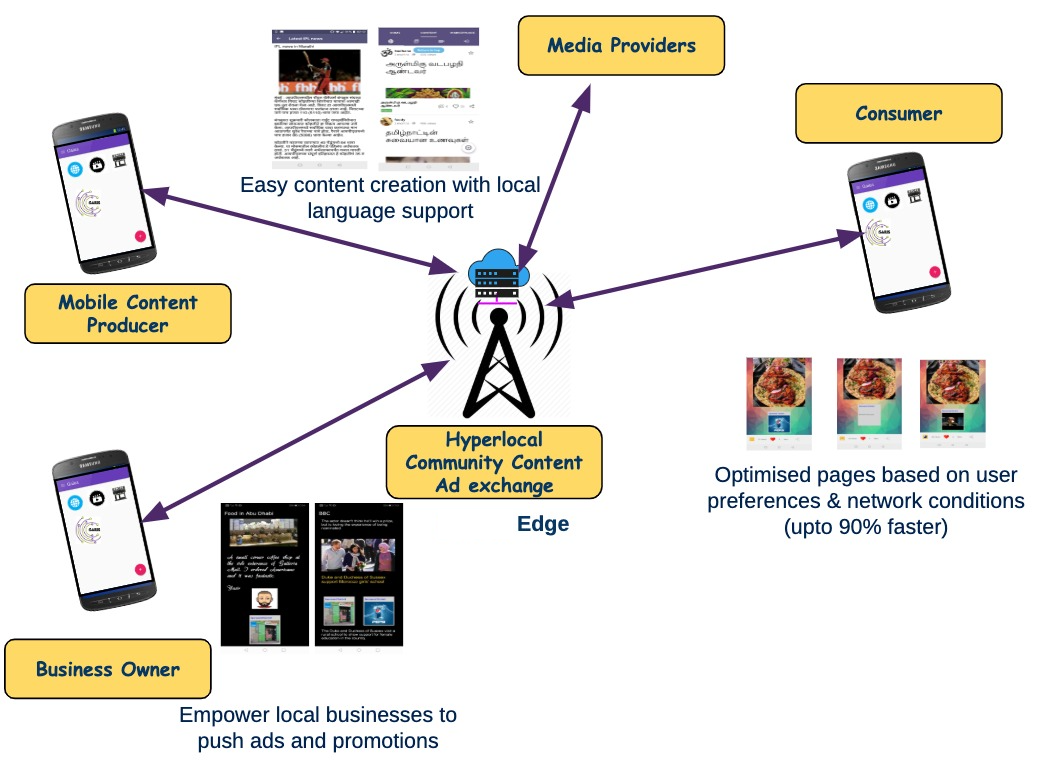}
  \caption{A high level view of the \name ecosystem}
  \label{gaius}
\end{figure} 

%The \name ecosystem solves the problem of content delivery from the hyper-local content/advertisement providers to mobile users. 
The main features of \name are: 
\begin{itemize}
	\item simple content creation: users can create content directly from their mobile phones using drag and drop
	\item light-weight content: the content is saved using our new Mobile App Markup Language (MAML), that eliminates today's web inefficiencies
	\item user controlled content delivery: our pages are generated on runtime, which ensures that each user gets a version of the page that matches his configurations and bandwidth limitation
	\item new economic ecosystem: through the content and local ads market place, where \name enables different tiers of local advertisements: small local individuals, small businesses, and large corporations
	%\item new way to identify and address content: \name content is divided into communities, where each is maintained by a single users or content provider.
\end{itemize}

The \name ecosystem as shown in Figure~\ref{gaius} comprises a set of distributed edge servers that help generate and serve different local pages at the edge based on user requests. Each edge is envisioned to serve a `region' (e.g., a city). One proposal for deployment is to place edge servers at a packet gateway level in a cellular core network. 
\subsection{\name Communities}
\begin{figure}[ht]
\centering
\begin{subfigure}{.3\textwidth}
  \centering
  \includegraphics[width=\textwidth, trim=0 70 0 170, clip]{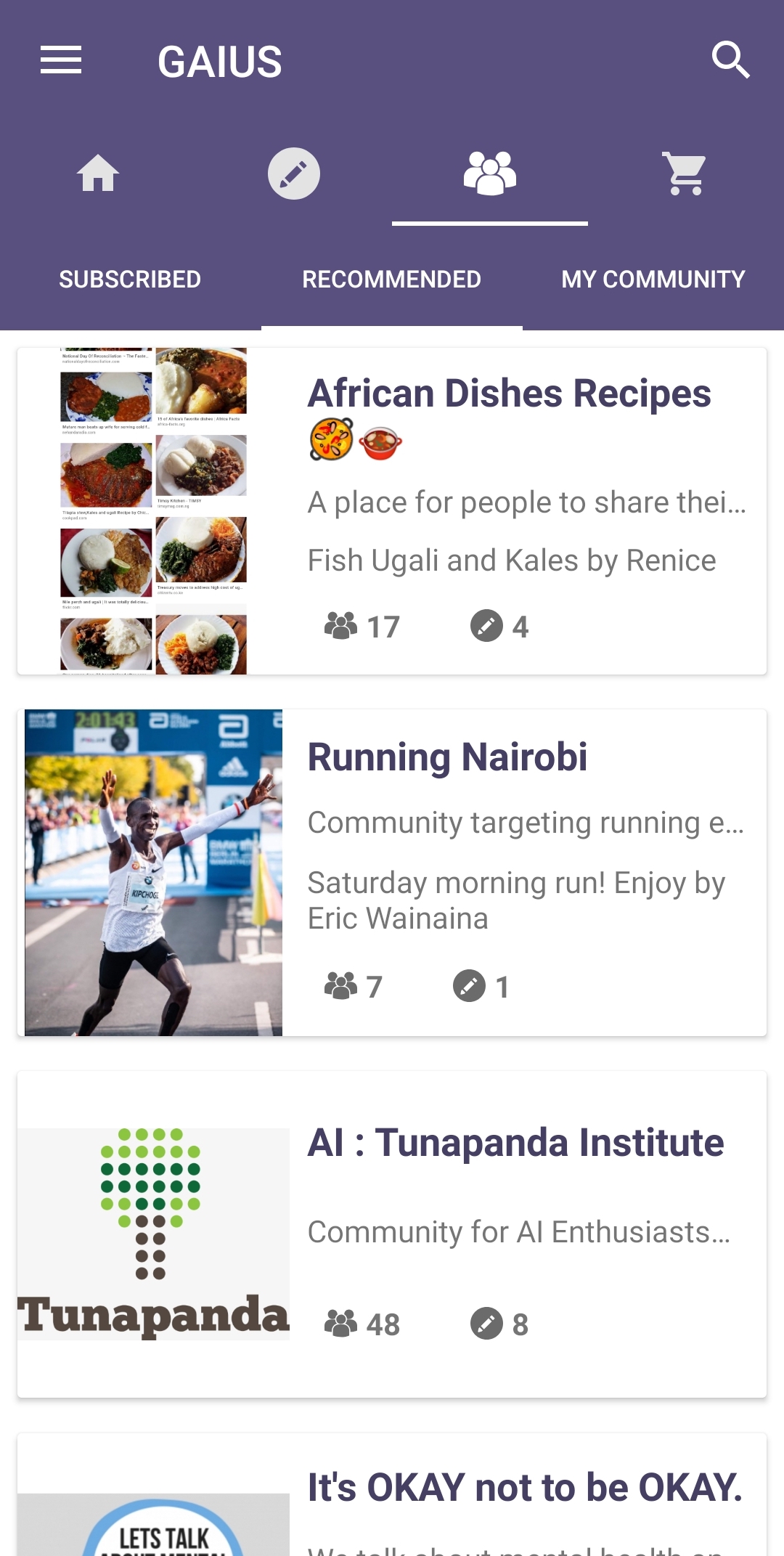}
  \caption{Kenya}
  \label{community1}
\end{subfigure}\hspace{0.01\textwidth}
\begin{subfigure}{.315\textwidth}
  \centering
  \includegraphics[width=\textwidth, trim=0 160 0 170, clip]{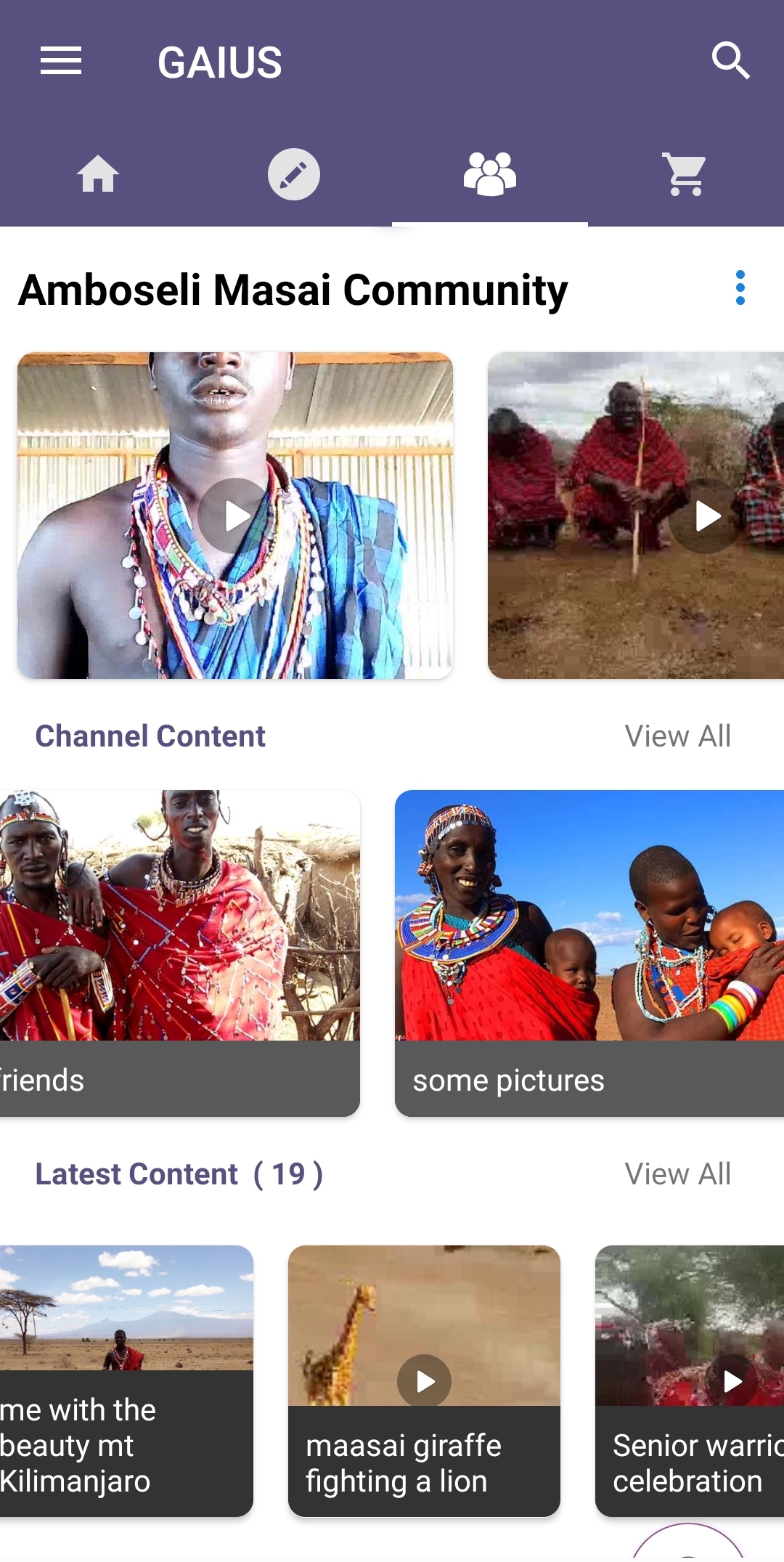}
  \caption{Amboseli Masai tribe}
  \label{community2}
\end{subfigure}\hspace{0.01\textwidth}
\begin{subfigure}{.3\textwidth}
  \centering
  \includegraphics[width=\textwidth, trim=0 70 0 170, clip]{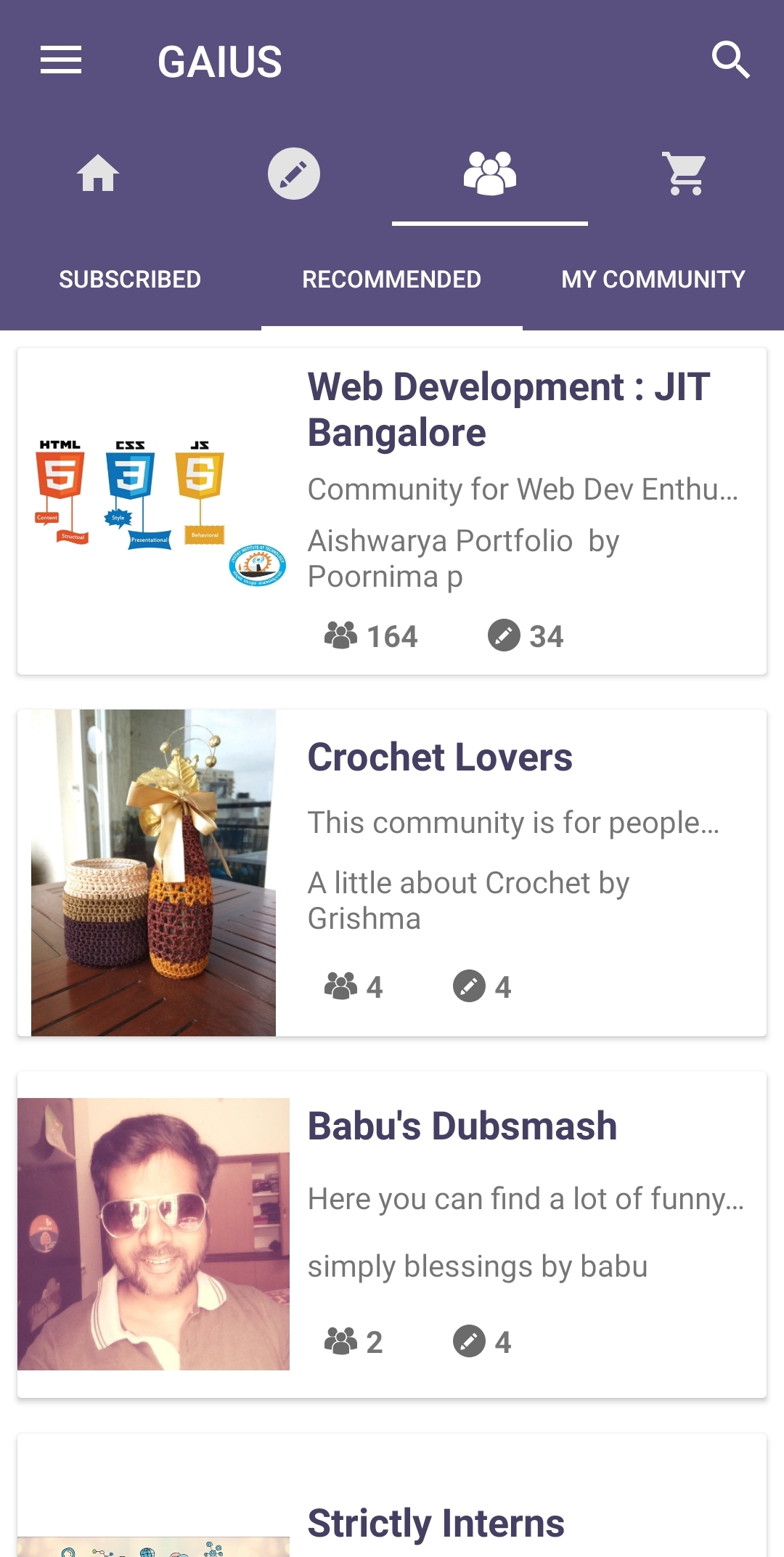}
  \caption{India }
  \label{community3}
\end{subfigure} %\hspace{0.02\textwidth}
% \begin{subfigure}{.23\textwidth}
%   \centering
%   \includegraphics[width=0.8\textwidth, trim=0 170 0 0, clip]{screenshots/indiaCommunity1}
%   \caption{Web development}
%   \label{community4} %\hspace{0.05\textwidth}
% \end{subfigure}%
\caption{\name example communities}
\label{fig:kenya_communities}
\end{figure}

%The legacy Internet relies on the notion of Uniform Resource Locators (URLs), where users type in the URL in a web browser to access the associated page. This requires the browser to do a DNS lookup to map the requested URL into the IP address of the machine that hosts the information. 
\name uses the notion of \textit{Communities} for users to access and search for content. A community is defined as a unique identifier for specific group of information that is hosted at the edge server. Each community is created by a unique user, content provider, or organization. A community is composed of multiple content postings as well as members. A community usually revolved around a common interest topic and users can join a community to easily view content from within that community. Figure~\ref{fig:kenya_communities} shows a couple of examples of communities in \name's deployments at Kenya and India. In addition, the figure also shows an example of how content is organized within a community (e.g. Amboseli Masai Community).

%This idea segregating the content into communities simplifies the user experience since users can join specific communities directly for content. Each edge server hosts hyper-local communities that are more relevant to the users in that locale. 

% , as well as the Web development community in Bangalore India.

%The local ecosystem takes input from different stakeholders (i.e. content providers, local advertisers, and users) and implements on-the-fly page creation. Figure~\ref{gaius} shows the inputs of the \name ecosystem. From these inputs, \name generates a web page and any associated advertisements for the user. The user's interface consists of a mobile application which enables user to connect to an edge and request web pages and communities. The \name app displays the web page using the MAML specification.

% \begin{figure*}[ht]
% \centering
% \begin{subfigure}{.23\textwidth}
%   \centering
%   \includegraphics[width=\textwidth]{screenshots/contentCreation1}
%   \caption{Content creation menu with different types}
%   \label{contentCreation}
% \end{subfigure}\hspace{0.02\textwidth}
% \begin{subfigure}{.23\textwidth}
%   \centering
%   \includegraphics[width=\textwidth]{screenshots/contentCreation2}
%   \caption{Simple webpage creation}
%   \label{webpageCreation}
% \end{subfigure}\hspace{0.02\textwidth}
% \begin{subfigure}{.23\textwidth}
%   \centering
%   \includegraphics[width=\textwidth]{screenshots/contentCreation3}
%   \caption{Simple webpage publishing}
%   \label{webpageSubmission}
% \end{subfigure}\hspace{0.02\textwidth}
% \begin{subfigure}{.23\textwidth}
%   \centering
%   \includegraphics[width=\textwidth]{screenshots/contentCreation4}
%   \caption{Simple blog creation}
%   \label{blogCreation}%\hspace{0.05\textwidth}
% \end{subfigure}%
% \caption{\name App simple content creation framework}
% \label{fig:simple_content_creation}
% \end{figure*}

\subsection{\name Simple Content Creation}
The \name framework allows uses to create any type of content from their smartphone. In today's web ecosystem, creating content and publishing it require certain tech skills: build a page natively from scratch, or use one of the existing Content Management Systems (CMS). The user also need to purchase a domain name, a hosting space, and possibly a way to advertise the page. All of these require certain level of technology awareness, as well as it incurs certain financial costs. This hinders the ability for users in developing regions to create and publish content.
% In addition, lots of today's CMS tools are only available for desktop machines and not for mobile.

\name is designed to facilitate faster and cheaper web access for users in developing regions to help share local knowledge and potentially spur economic growth. 
% There are many possibilities for locally relevant content that is especially important in a specific community. For example, information related to health, how to fight or prevent certain diseases, agricultural information, upcoming local events, trading and sales information, services offered by community members, etc. 
We envision local content being created by three main sources: individual community members, organizations operating within these communities (e.g., religious institutions, UNICEF, local NGOs etc.), and local content providers (e.g., a local news site). The usefulness of local content is not confined to the content itself, but also its form. There may be, for example, language and dialectal barriers, where community users are interested in accessing information written in a local language. Some communities may have users who are illiterate who would be better served by largely visual content. 
% While these accessibility and localization-related aspects do exist in various diffuse forms on today's Internet, they are hard to find and incoherent.

In \name, we have simplified the notion of content creation by building an in app content creation framework. Users can simply create any kind of content easily from their smartphone using the app. 
% Figure~\ref{fig:simple_content_creation} shows the menu for the content creation, as well as different examples on how web pages or blogs are created in \name. The figure shows that 
The app content creation has a simple menu design that consists of several basic elements for building a page. The user experience is very intuitive, where by drag and drop the user can move things around, resize elements by pinching in and out. Once the page is finished the user can publish it with a click of a button.
% by filling a simple form that asks the user to fill the page caption and giving a couple of relevant tags that associate with the page content. 
The app automatically fill in the used language and the location, these are used to filter content based on their distance to the user as well as the user selected language. Once the content is published, all users within the community are able find it. Since all of the local content is hosted nearby, the hosting and bandwidth costs can be minimized. \name relies on the local ad marketplace to cover the costs of local content. 
% The template provided by \name also has place holders for displaying local ads to users browsing these local communities. 

\subsection{Mobile App Markup Language (MAML)}
% \begin{comment}
    
% To overcome the web page complexity, 
The MAML web specification is designed to provide a flattened page structure that eliminates the need for recursive object calls, thus reducing the overall page load times. MAML simplifies the way in which styling and interactivity is specified, the main page consists of all relevant objects that are needed to represent the overall look and feel of the page. Furthermore, each object specifies all necessary information about itself in relation to the page: relative location, dimensions, styling information, the URL of the object's source, and hyperlinks. By flattening the page representations, the only additional web requests are to fetch objects that do not already render on the page (e.g., audio or video). In regular HTML pages, the complexity of rendering a web page after the downloading of the first \textit{index.html} is dictated by the different DNS resolutions, redirects and inherent complexity in the object dependency tree. 
% embedded in the web page. 
% In contrast, MAML removes this complexity by simplifying the way a web page is written. This makes the rendering of the MAML pages extremely simple and enables prioritization of object downloads.
To retain functional compatibility, MAML pages must be similar in look and feel as conventional HTML web pages. Consequently, the current MAML specification supports six different object types: \textit{image}, \textit{text}, \textit{video}, \textit{rectangle}, \textit{text-field}, and \textit{button}. For example, the MAML image object represents the mapped HTML image tag, which contains the URL of the actual image that needs to be prefetched from the network. The image object has the current attributes: URL, x and y coordinates, width and height of the image, and an optional hyperlink in case the image is clickable. An example of such an image object is given in Table~\ref{tab:table1}, which represents a logo image with its relevant location and size information. 
% Additionally, the table also shows examples of the other types of MAML objects.
The MAML representation of a page provides several important benefits for bandwidth constrained users:

\begin{enumerate}
	\item MAML object specification incorporates styling and interactivity, thus eliminating the need for JavaScript and CSS. This helps cut down the size of the page and reduces page load time. 
	\item Upfront specification of objects in a page enables prioritization of downloads. 
	\item The flattened tree structure simplifies the object loading, object parsing, and the object rendering, thus enabling faster page rendering.
\end{enumerate}
\begin{table}[hbt]
	\begin{center}
	\begin{tabular}{ |c| p{6cm} | } 
	\hline
	\multicolumn{1}{|c|}{\textbf{Object}} & \multicolumn{1}{c|}{\textbf{Example}} \\\hline
	image & \textit{\{"type":"img","url":"www.example.com/} \\
	 	  & \textit{logo.png","x":43.0,"y":57.0,"w":390,"h":60\}} \\\hline
	text & \textit{\{"type":"txt","txt":"Example text of page",} \\
	 	  & \textit{"x":65.0,"y":867.0,"w":950,"h":31,"font":20,} \\
	 	  & \textit{"font-type":"Arial","color":"\#946c3b"\}} \\\hline
	rect & \textit{\{"type":"rect","x":0,"y":28,"w":1080,} \\
	 	  & \textit{"h":147,"color":"\#ffffff"\}} \\\hline
	video & \textit{\{"type":"video","url":"www.example.com/} \\
		  & \textit{video.mp4","x":82.0,"y":60.0,"w":626,"h":352\}}\\\hline
	\end{tabular}
	\caption {Examples of different MAML object types} \label{tab:table1} 
	\end{center}
\end{table}

\noindent\textbf{RSS Feed Parser:}
Various RSS feeds were translated into the MAML format using an RSS to MAML translator. We have chosen more than 50 highly popular content feeds from within the cities of deployment. The RSS Feed Parser was written in python using the "feedparser" module. Each individual item within the RSS feed was mapped onto three different MAML objects: an image object that contains the RSS item image, and two text objects, one for the title and the other for the description. The RSS item hyper-reference was added to both the image and the title text object, so that users can click on an individual item and open the full article on the web. The RSS feeds parser is configured to automatically update the MAML pages every 30 minutes. 
% One of the issues that we faced when translating RSS feeds was that different content providers have different ways of adding content within the XML format. For example, the image associated with some content is added using different tags within each RSS feed XML and %,  such as \textit{media\_content}, \textit{thumbnailimage}, \textit{img}, or \textit{media\_thumbnail}. 
% In other cases the image was embedded within the content description. We custom tailored the translator to deal with these different situations, and to be able to download the image from the URL.
% We also saved two additional versions of each image: one for the medium fidelity setting, and one for the low fidelity setting. 
% One issue that we had to deal with is identifying the MAML text object height, as this changes based upon the size of the text and the font size. We used the freetype library in order to convert each character to a glyph index and then compute the width and height of each character. Then based on the predefined MAML text object width we computed the height of the entire paragraph.

% \end{comment}

\subsection{Data Saver and Content Policy}
One of the motivations behind the \name ecosystem is to serve users different versions of the same content dictated by policy inputs. For example, based on the user budget, he/she can request a version of a page that is optimized for lower data consumption. However, user-side economics is not the sole determining factor, a number of additional parameters also play a role in page creation. The page layout and the content dependencies are typically determined by the content provider and the priority of the content. Hence, content providers specify the overall layout of the page as well as how the content is related and structured. On the other hand, advertisers determine which ads needs to be displayed based on their own economic considerations (e.g., relevance to users, revenue, advertisers budget). For fast page loads, the users' network conditions also play a role. All of these factors are combined to produce the final page. 

\subsection{\name Advertisement Ecosystem}
Online advertising plays a critical role in today's Internet ecosystem. Our vision for \name as a localized content platform extends to advertisements as well. \name aims to enable a local advertisement marketplace that provides a platform for local advertisers to push advertisements to their communities. Each \name edge constitutes a local marketplace where local and global advertisers can compete.  The \name application allows users to easily publish their own advertisements in form of images or text from their phone. The ad is then added to the local ad market exchange to be displayed to users within their community.
%Local content and advertisements are extremely important within the context of developing countries because most web content produced today targets users from developed regions where good Internet connectivity is readily available at very little cost.  

%\name targets the local ad market through a three tier ad model. The first tier focuses on individual users whom might want to advertise their own small products or businesses. The second tier focuses on small local advertisement agencies that usually use local magazines or newspapers to advertise. The third tier includes big corporations that advertise on larger scales. \name supports all three tiers, but we initially focusing on the first two tiers. The \name application allows users to easily publish their own advertisements in form of images or text from their phone. The ad is then added to the local ad market exchange to be displayed to users within their community.

Within \name, advertisers can specify their targets and costs roughly analogous to advertising today on Google or Facebook. In \name the advertiser specify rules for:
\begin{itemize}
\item Visibility of the ads within the local community.
\item Accessibility beyond the local community.
\item Targeting advertisements for audiences in local and non-local contexts.
\end{itemize}

These rules enable both global and local content providers to specify their advertising policy. A local content provider, for example, can choose to only show their advertisements to users connecting from a particular geographic location. Advertisers are currently charged using a fixed model of revenue based on how many users they want to reach in a particular \name edge.

\name is a collection of highly localized ecosystems where each ecosystem has 100-100k users, so our ad revenue model is completely different from conventional content providers. In \name we implement the aggregator logic as part of our edge logic. At the edge, ads are chosen based on location, content providers input, and users interests. Unlike conventional aggregators that collect a fixed fraction of ad revenue generated~\cite{Gill:2013:BPF:2504730.2504768}, our aggregator collects a variable fraction based on the network infrastructure maintenance costs. In essence since the network infrastructure cost remains fixed for a localized community, the ads become cheaper over time as more content providers start using the system. In this manner, \name supports the local content ecosystem. Our overall pricing model is also simpler, i.e. advertisers are charged weekly based on the base price of an impression (paid to the content provider) and network infrastructure maintenance cost. This means that \name enables cheaper ads which may have a larger relevance per impression. We envision this model will enable small local businesses to advertise within \name, increasing the number of people who have access to marketplace. 
% A detailed economics modelling comparison between \name and conventional content providers remains part of our future work. 

%\subsection{\name Marketplace}
%\name also provides a hyper-local marketplace for users to advertise items that are for sale, as well as for businesses to advertise their products. Again through a simple interface \name users can easily create a market posting and add relevant information for potential buyers/customers such as their contact phone numbers as well as their locations. This information is then linked to the market posting, where users can simply click the map to be forwarded to Google maps. We also link the post to Whatsapp if the potential buyer would like to chat to the owner of the post. Figure~\ref{fig:gaius_market} shows examples of the marketplace.

%\begin{figure}[ht]
%\centering
%\begin{subfigure}{.21\textwidth}
 % \centering
  %\includegraphics[width=0.7\textwidth, trim=0 550 0 0, clip]{screenshots/market1}
  %\caption{Example from the Kenyan market place}
  %\label{market1}
%\end{subfigure}\hspace{0.05\textwidth}
% \begin{subfigure}{.2\textwidth}
%   \centering
%   \includegraphics[width=\textwidth]{screenshots/market2}
%   \caption{Example from Indian market place}
%   \label{market2}
% \end{subfigure}\hspace{0.05\textwidth}
%\begin{subfigure}{.21\textwidth}
 % \centering
  %\includegraphics[width=0.7\textwidth, trim=0 0 0 550, clip]{screenshots/market3}
  %\caption{Example of an item in the market place}
  %\label{market3}
%\end{subfigure}
%\caption{\name Market place}
%\label{fig:gaius_market}
%\end{figure}

%% file: implementation.tex
\section{Implementation}

% The \name implementation is divided into two main parts: an edge server, and an Android mobile app. 
% % In addition, we have also implemented a Desktop PC tool to create, render, and visually edit MAML pages. 
% In this section we detail each of these components.

\noindent\textbf{\name Edge Server:}
The \name edge server is based on Apache Server, using the the Common Gateway Interface (CGI). When a specific user request a MAML index file, Apache redirects the request to our \name policy engine. The user request specifies the fidelity level that the user has specified. Using the fidelity level, the policy engine creates and delivers a custom version of the page on the fly. In addition, the \name ad selection policy mixes the local content with relevant ads from the ad market place. The ad selection policy chooses ads based on the fidelity level and the relevance of the ad to the user. A high fidelity chooses ads with high quality video, whereas a medium fidelity chooses an image quality ads, and a low fidelity chooses textual ads from the ad marketplace.
% The edge server maintains a list of the information channels it hosts. This is used to answer user queries about the available information channels. Based on a user's request, a specific version that matches the user preferences is served to the user.
% \begin{figure}[hbt]
%     \centering
%     \begin{subfigure}[t]{0.21\textwidth}
%         \centering
%         \includegraphics[width=\textwidth, trim=0 450 0 0, clip]{figures/cnn_hi2.png}
%         \caption{High resolution (377 kB)}
%     \end{subfigure}%
%     ~
%     \begin{subfigure}[t]{0.21\textwidth}
%         \centering
%         \includegraphics[width=\textwidth, trim=0 450 0 0, clip]{figures/cnn_low2.png}
%         \caption{Low resolution (73 kB)}
%     \end{subfigure}
%     \caption{\name rendered CNN web page with different user fidelities}
%     \label{komodo_results}
% \end{figure}
We emulate content providers by creating multiple versions of several popular websites, where each version contains different image sizes by manipulating their resolution. For example, for the CNN web page a high fidelity version of the page has a page size of 377 kB, whereas a low fidelity version has only 73 kB size. This means that the lower fidelity is significantly smaller for the lower resolution version (80\% reduction in size).
% Figure~\ref{komodo_results} shows two versions of the CNN web page with different fidelity levels (high vs low resolution). 

\noindent\textbf{\name Android App:}
The app has a simple user interface with multiple features: content browsing, content creation/editing, and ad creation. The content browsing feature displays the different content to the user sorted based on a mixture of the content proximity to the user and the content popularity. 
% The list displays the channel title and icon for each information channel. 
When the user clicks on a specific content, 
% the app requests the corresponding information channel. This can lead to one of the following: 
if that content is a page then the corresponding MAML page is sent to the user and is rendered by the app browser. 
% on the other hand, if the information channel consists of several sub-pages then a list of sub-pages is sent to the app from the edge server and is displayed to the user to choose from. 
The \name app user is able to set his own fidelity level preference: high, medium and low. Upon setting the fidelity, the corresponding image quality from the edge server changes. In addition, the edge server changes the type of the ads within the requested MAML pages, from video ads at the high fidelity level, images at the medium fidelity, and plain text ads at the low fidelity.

%% file: eval.tex
\section{\name Performance}

{\blue We evaluated the performance of the \name ecosystem on three fronts. First, we evaluate the performance gains achieved by using the simplified MAML web specification instead of HTML on 30 popular webpages. We show that, under the same network conditions, this specification tends to significantly reduce the page complexity; This allows time taken for the MAML-converted webpages to load to be significantly lower than the time taken by the original versions across our experiments. Secondly, we compare the performance of the \name application using the \name edge server against two common web browsers and show that the same webpages need less time to load for \name due to the close proximity to the edge server, even in cases where the page size is similar for example, when the OperaMini optimized and compresses webpages are compared against \name webpages requested at high fidelity. Third, we compare the performance difference in the different fidelity levels of and measure the improvements in page loading times for each configuration. The current evaluation is limited in scope and is intended as a proof of concept to illustrate the potential improvements offered by the \name ecosystem as opposed to a comprehensive analysis of its usability and impact in each locality where it is deployed using detailed usage statistics for individual users of the \name ecosystem.}

\subsection{MAML Evaluation}
% \begin{comment}
% MAML pages are designed to simplify the complexity of web pages and enhance performance in developing regions without significantly altering the feel of the pages. 
We evaluate performance gains with respect to three common web performance metrics: page load time (PLT), overall page size, and the total number of requests~(for objects) generated by a page. We compare these metrics against those of the original HTML pages downloaded through a normal browser. We compared 30 different web pages taken from Majestic Million's~\cite{majestic} top 100 pages. First, we request these 30 HTML pages using the Selenium Webdriver and save the HTTP Archive (HAR) files. HAR files log page load details including the number of objects in the page, time taken to download each object, and total page load time. We then convert the same 30 web pages using the new MAML page. 
% We compare the page load times of the original HTML pages against those of the MAML pages on a desktop PC. %JAY: specs?

\begin{figure}[hbt]
\centering
\begin{subfigure}{.35\textwidth}
  \centering
  \includegraphics[width=\textwidth]{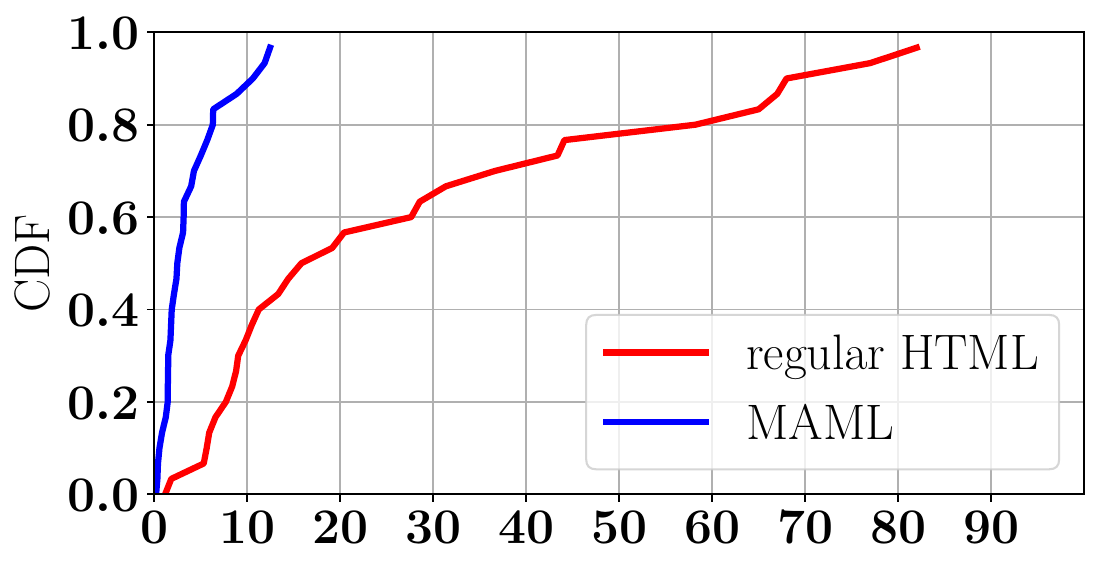}
  \caption{Page load time (s)}~\label{fig:plt}
\end{subfigure}%
\begin{subfigure}{.35\textwidth}
  \centering
  \includegraphics[width=\textwidth]{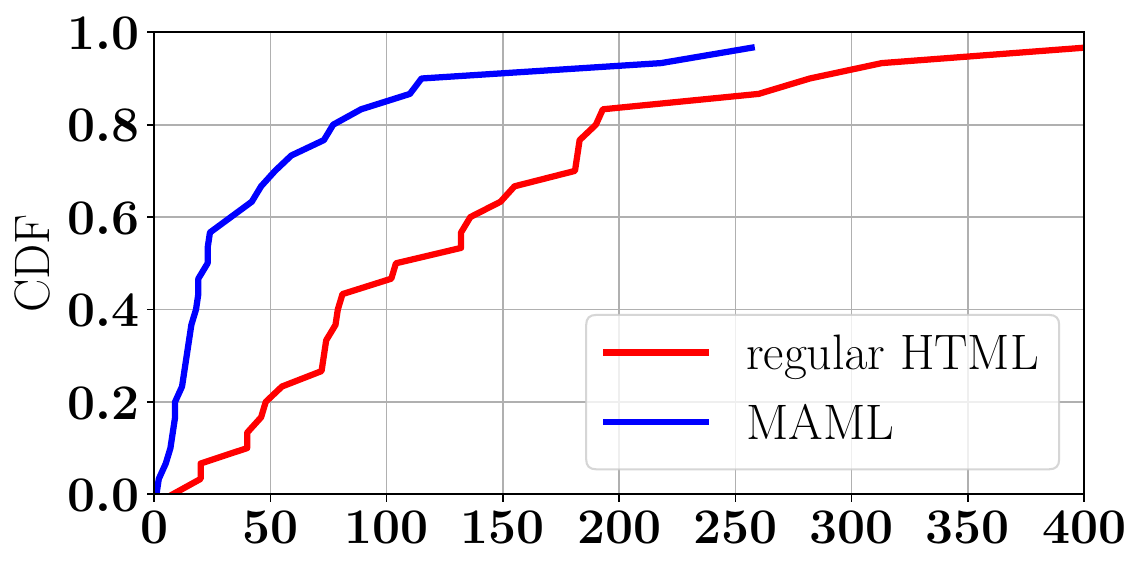}
  \caption{Per page requests (\#)}~\label{fig:numReq}\\
\end{subfigure}
\begin{subfigure}{.35\textwidth}
  \centering
  \includegraphics[width=\textwidth]{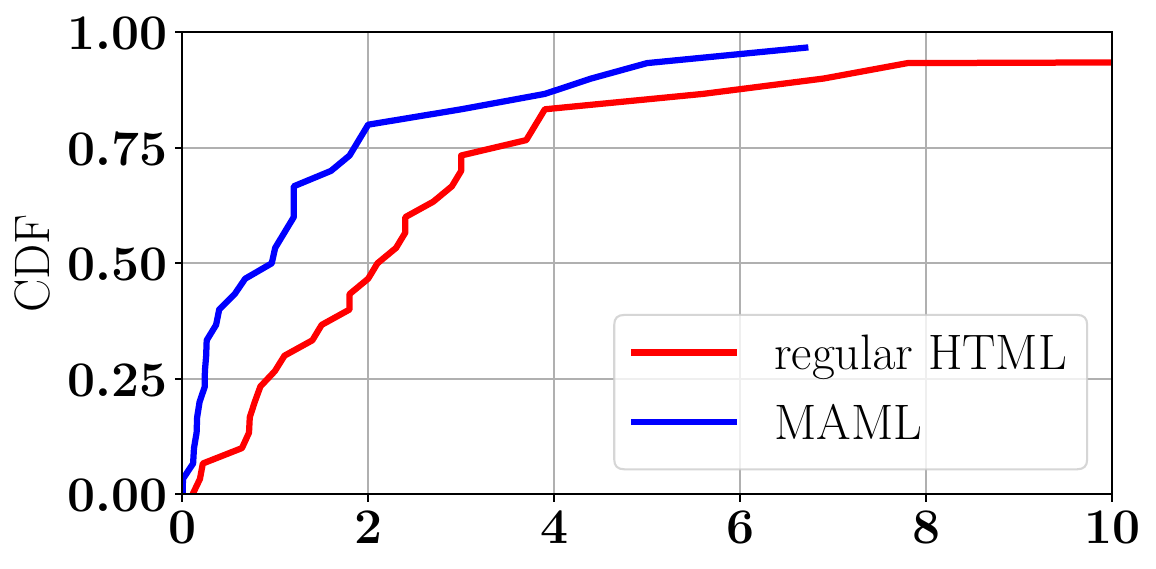}
  \caption{Page size (MB)}~\label{fig:pageSize}
\end{subfigure}%
\begin{subfigure}{.35\textwidth}
  \centering
  \includegraphics[width=\textwidth]{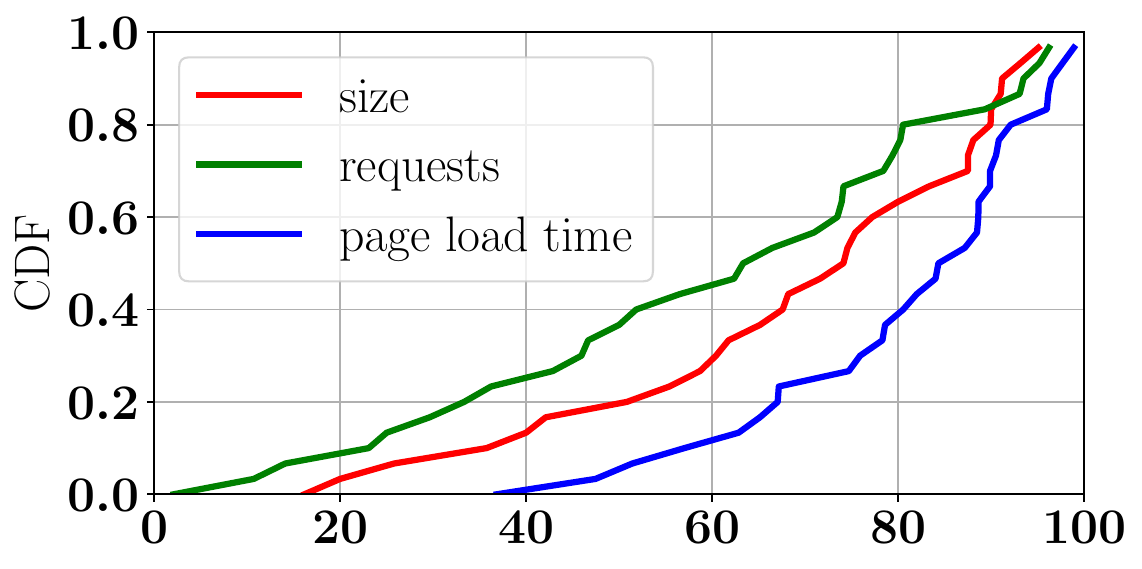}
  \caption{MAML reductions (\%)}~\label{fig:mwtGains}
\end{subfigure}
\caption{MAML page vs.\ regular HTML}
\label{fig:testing}
\end{figure}

Figures~\ref{fig:plt} shows the cumulative distribution functions (CDF) of the PLTs comparing the MAML pages with the HTML pages. The blue curve shows the MAML PLT, whereas the red curve shows the HTML PLT. The result shows that the MAML pages load significantly faster than the regular HTML pages, achieving a median PLT reduction of more than 80\%. The figure also shows that in the most extreme case the MAML PLT does not exceed more than 12 seconds, compared to more than an 80 seconds PLT for the regular HTML page.

Figure~\ref{fig:pageSize} show the CDFs of the page size for both MAML page and HTML. The results show that MAML page is 60\% smaller in size than an HTML page (by comparing the median value of the two curves). The page size reduction is because MAML pages cut down all extra CSS, JS, and HTML objects. Similarly, Figure~\ref{fig:numReq} shows the CDFs comparing the number of object requests per page. Again MAML page manages to reduce the number of requests by about 80\%.
Figure~\ref{fig:mwtGains} shows the CDFs of the reduction percentages achieved by MAML over HTML across the three metrics of interest. We can observe that MAML page reduces the median values of all three metrics by 60\% - 80\%.

\subsection{\name Edge Evaluation}
To evaluate the performance of the edge server, we translated a few hand-chosen web pages from Alexa's top 100 sites from different developing countries. We hosted these pages at a \name edge server and compare the performance of \name against two mobile web browsers: OperaMini and Google Chrome. We used the WebPagetest web performance tool~\cite{webpagetest} to obtain the performance results of OperaMini and Google Chrome. Our setup consisted of a Samsung Galaxy S4 Android phone, which is connected to our WebPagetest hosting machine. We requested each of the selected Alexa web sites 18 times for OperaMini, Google Chrome, and the \name app. We measured the PLT and page size of the selected webpage. For \name we requested the pages with three different fidelity levels (high, medium, and low) to simulate different network conditions and user preferences. 
% Low fidelity represents a user with poor connectivity, while high fidelity represents a user with good connectivity. 
\begin{figure}[hbt]
\centering
\begin{subfigure}{.35\textwidth}
  \centering
  \includegraphics[width=\textwidth]{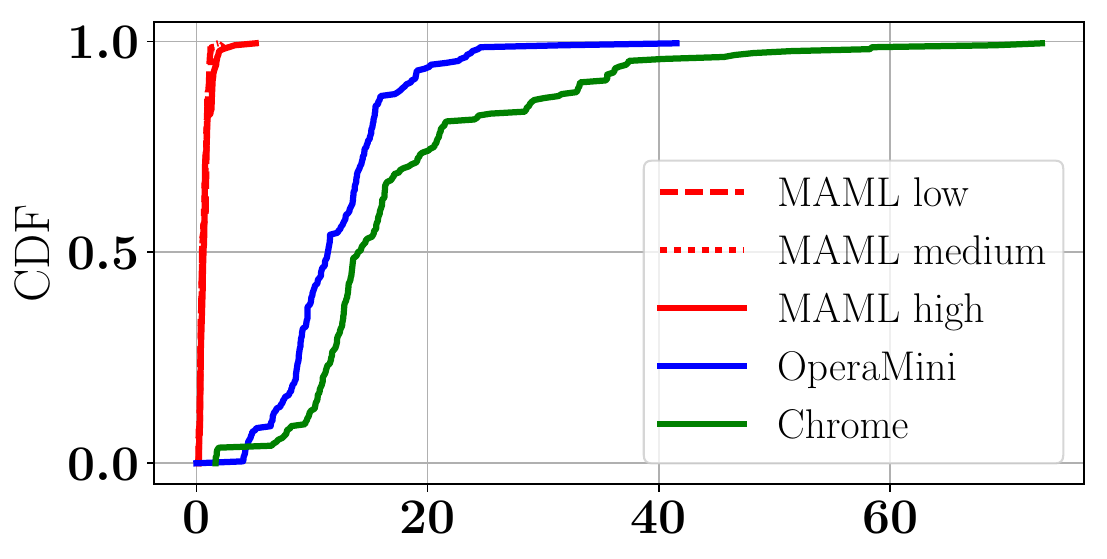}
  \caption{Page load time (ms)}~\label{fig:mobplt}
  \label{fig:test1}
\end{subfigure}%
~
\begin{subfigure}{.37\textwidth}
  \centering
  \includegraphics[width=\textwidth]{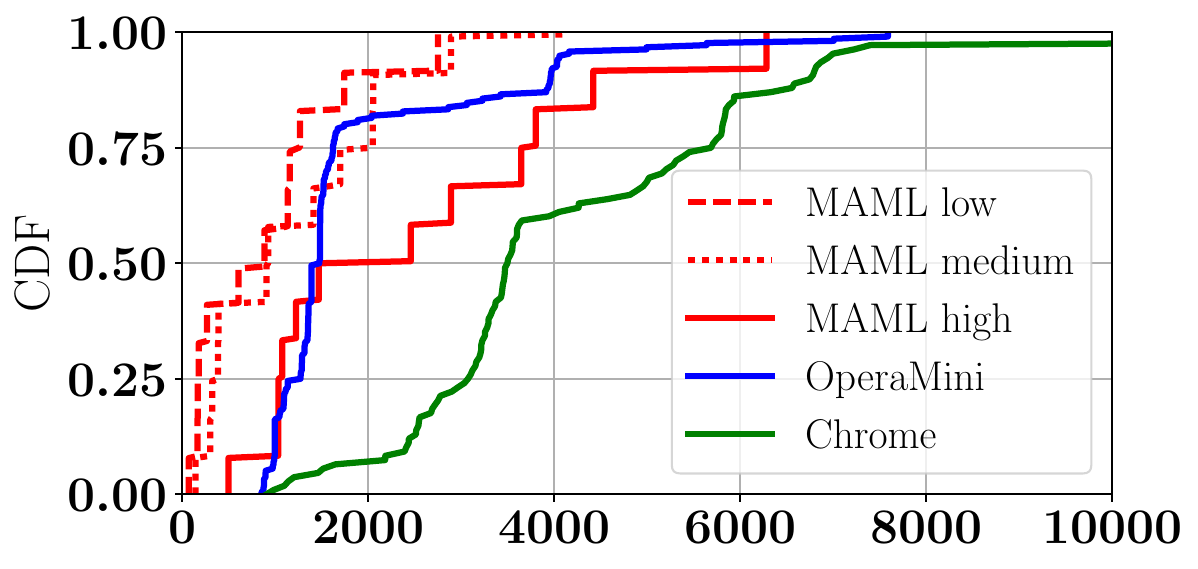}
  \caption{Page size (kB)}~\label{fig:mobpagesize}
  \label{fig:test2}
\end{subfigure}
\caption{\name Comparison to OperaMini and Chrome on Mobile}
\label{fig:test}
\end{figure}

Figure~\ref{fig:test1} shows the CDF of the PLTs for all three browsers. We can observe that Chrome has the highest PLT compared to OperaMini and \name. %This is due to the fact that both OperaMini and \name were designed to take certain web optimization techniques into consideration that enhance the performance. 
OperaMini had a lower PLT because it requests the pages through dedicated servers that collect all of the necessary resources of the page, optimizes, and compresses the page before sending it to the client. In contrast, \name already hosts the MAML representation of the pages on an edge server that is close to the client, which enhances the PLT several-fold. The second important metric that we compare is the overall page size. The size of a page is critical in developing contexts because many people are constrained by a pay-as-you-go data subscription model, or they have relatively small data plans. 
% Thus, being able to optimize and reduce the page size would significantly reduce their data consumption and browsing cost.
Figure~\ref{fig:test2} shows the page size CDFs comparisons again for all three browsers. We can observe that Chrome has the largest page size. This is unsurprising given that Chrome does not do any optimizations to reduce the page size. If we compare OperaMini to \name MAML's high fidelity, we can see that in 50\% of the cases, \name serves slightly smaller pages compared to OperaMini. However, for the remaining 50\%, OperaMini does perform better in terms of page size. If \name operates at lower fidelity settings (dotted and dashed red curves), we can observe that in comparison to OperaMini, \name reduces the median page size by 57\% and 35\% at the low and medium fidelity levels, respectively.

\subsection{Page Load Times and Sizes}
\begin{figure}[hbt]
\centering
\begin{subfigure}{.4\textwidth}
  \centering
  \includegraphics[width=\textwidth]{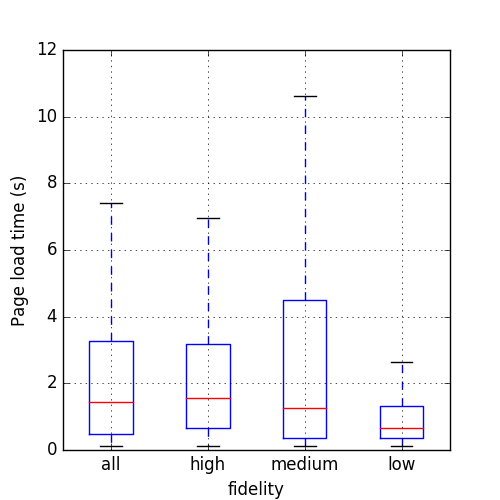}
  \caption{\name PLT}
  \label{fig:gaiusplt}
\end{subfigure}%
~
\begin{subfigure}{.44\textwidth}
  \centering
  \includegraphics[width=\textwidth]{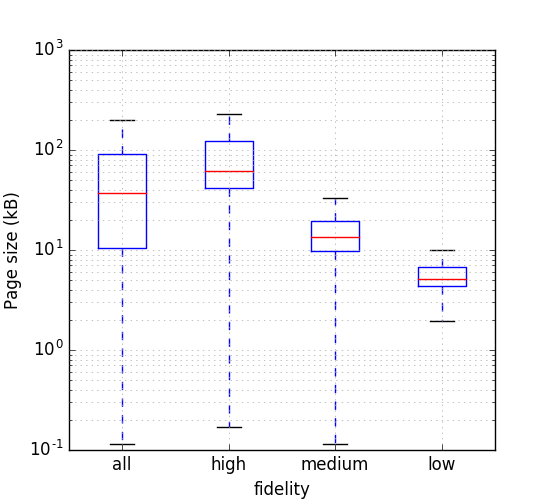}
  \caption{\name page size}
  \label{fig:gaiuspagesize}
\end{subfigure}
% \begin{subfigure}{.23\textwidth}
%   \centering
%   \includegraphics[width=\textwidth]{figures/plt_network}
%   \caption{Box plot showing the distribution of page load times for different network types.}
%   \label{fig:gaiuspltnetwork}
% \end{subfigure}%
\caption{Box plot showing the distribution of page load times and page size for different fidelity levels. }
\end{figure}

To evaluate the current deployment of the \name, we record statistics for each individual page request: the overall page load time, the overall page size, the page fidelity requested by the user, the user's geographic-location, the user's network type and mobile phone model. Figure~\ref{fig:gaiusplt} shows the PLT box plot for the overall page requests as well as for the different individual fidelity levels. The x-axis shows the fidelity level, whereas the y-axis displays the PLT. It can be seen that the median PLT is about 1.6 seconds, with the 75\% not exceeding 3 seconds. In addition, it can be seen that the "high" fidelity level has the highest median PLT, in contrast to the "medium" fidelity which has a lower median value, and the "low" fidelity having a median PLT below one second. Figure~\ref{fig:gaiuspagesize} shows the page size box plot again overall and across the multiple fidelity levels. The figure clearly shows how effective \name is in reducing the overall page size with a median smaller than 40kB. The figure also shows how the medium and low fidelity manages to reduce the overall page load time by more than 6x.

%% file: deployment.tex
\section{\name Experiences}
We discuss our deployment experiences from July 2019 to January 2020. We deployed \name in several communities across three countries: Kenya, India and Bangladesh. The deployments consisted of a single \name server per country and a number of smart-phone users from different communities. These servers were used to locally host the content, as well as all our back-end APIs and database. To ensure initial user stickiness, we pre-populated our server with local content specific to user locations so that users have relevant content to consume when they log into the \name app. This initial content was created by translating popular websites and RSS feeds within the users' city. The content was chosen from a wide variety of categories ranging from news, sports, entertainment, events, food, etc. 

\subsection{Kenya}
Our user acquisition followed a partnership approach for user acquisition and teamed up with a non-profit social enterprise that runs intensive 3-month technology, design, and business training courses in extreme low-income environments of East Africa such as Kibera (the second largest urban slum in the world). The enterprise helped organize workshops to teach users in Kibera how to install and use the app. They have also helped in organizing street visits to recruit small business to promote their content within the \name ecosystem. 

The users in Kenya interacted the most with \name's market place with over 200 different businesses posting in the Kenyan market place ranging from scrap metals, second hand clothes, custom furniture, to fresh fruits and dairy products. Users in Kenya used \name to create several interesting communities that stems from the needs and hobbies of people there (e.g gaming). %For example there were a couple of hair related communities. 
% One of the highly populated communities was 'Kenya Gaming Community' with over 100 users posting about different styles of games from computer games to even simple board games. 
The most interesting community in Kenya was the {\it Amboseli Masai Community} (Figure~\ref{community2}), belonging to the Amboseli Masai tribe. They used \name to publish different content related to the tribe culture and festivities. They showcase different video content on different skills such as: building beaded bracelets and necklaces, cultural dances,warriors celebration, fire making, etc. The Masai tribe is currently using the community as a platform to spread their culture and showcase different aspects of their life.

\subsection{India}
Our user acquisition focused on three cities: Bangalore, Chennai and Mumbai (and surrounding areas). We carried out a mix of user acquisition strategies: working with local organisations such as community radios, recruiting small businesses through door to door visits, arranging several workshops in colleges across these cities as well as running targeted Google Ad campaigns. In this process, we partnered with 11 community radios in Mumbai (and surrounding areas) that streamed radio programs on demand via our platform, recruited more than 100 local businesses ranging from small tea shops, outlet factories, restaurants, food trucks and acquired 1000+ students across Bangalore and Chennai. 

A number of educational communities offering multiple educational content were also created around different colleges in both Bangalore and Chennai. These communities offered a number of classes such as web development and python programming to students in these colleges. Instructors used \name's communities to teach, share course materials and engage with the students. Businesses around these colleges utilised the \name platform to offer vouchers and discounts to the college students. 

\subsection{Bangladesh}
The user acquisition strategy that we used in Bangladesh was Google Ad campaigns. Unlike in Kenya, Google Ad campaigns were extremely successful, in fact we managed to acquire more than 80,000 users. Bangladesh was one of our most successful deployment where people were eager to use the platform mainly for entertainment purposes. The active users pushed multiple short clips, videos, memes, and images. The users have also created multiple communities, especially for local celebrities sharing their latest news and images.

\subsection{\name App Reviews}
We have had more than 100,000+ users downloading and using the app with 317 playstore reviews giving a 4.4 star rating.  A number of the reviewers expressed the simple nature of the app, where one of the reviewers has mentioned: \textit{``It's self explanatory and easy to use. A very fantastic app that incorporates nearly everything''}. Another reviewer mentioned: \textit{``This app let anyone to create a mobile webpage for free with a few clicks from your phone itself. Excellent app all around''}.

Other reviewers were happy with the business aspects provided by the app, one reviewer from India said: \textit{``This is the best app I have on my phone. Creating mobile webpages for my business has never been so easy thanks for the \name app. I also read all my local news from Chennai via the app and get all the local offers nearby. Thanks \name team for creating such a fabulous app''}. 

Finally, two reviewers talked about the community and business adverts by saying: \textit{``I love the community aspect. The app is a great platform''}, and \textit{``Its not slow works very well and has everything from business adverts to informative news''}.

%% file: conclusion.tex
\section{Discussion and Lessons Learnt}
Based on our experiences deploying \name in several emerging markets, we outline some of the key lessons learnt.

{\bf Need for a Hyperlocal Web abstraction:} Across all our country deployments, we observed that users appreciated the concise hyperlocal Web abstraction offered by the \name app. When presented with different forms of content, we observed that users interacted more with locally relevant content than global content on the platform. The \name App enabled users to personalize their choices as a function of content topics, community interests and marketplace interests. We also observed that a significant fraction of users were pure content consumers compared to the number of active content curators.

{\bf Purpose of Hyperlocal Communities:} The nature and purpose of communities that were formed on the platform varied a lot by country even though we attempted to popularize similar communities across geographies. In India, we found educational communities to find the strongest use-case for the platform while we had larger marketplace adoption among small businesses in Kenya. Bangladesh users found a larger use around entertainment centric content.

{\bf Economic Challenges in Emerging Markets:} {\blue The \name platform was well-received in the communities where it was deployed, amassing a significant userbase of more than 100,000 people downloading and using the platform. Gaining enough active local users to develop a large ecosystem proved a difficult task to achieve and required aggressive marketing strategies and a massive investment of capital and effort. Despite this reception, our deployment of the \name ecosystem in emerging markets was met with economic challenges that hindered the sustainability of such a platform in developing regions. These challenges included significant expenses incurred in the deployment and maintenance of the cloud network infrastructure required to support the \name ecosystem, which includes application servers, load balancing, and cloud storage. In comparison, the revenue generated by advertisements on \name was limited due to factors such as low click-through rates and limited production and dissemination of hyperlocal advertisements. 

In order for such a platform to be sustainable in developing regions, the differences in infrastructure costs in developing regions need to be proportionally lower to the decrease in revenue generated, however, the cloud network infrastructure costs in developing regions do not maintain this property, as corroborated by a comparison of the cloud infrastructure costs for Amazon Web Services obtained using the AWS pricing calculator~\cite{awscalc}. Additionally, the amount of ad content in a region needs to exceed a certain threshold to generate enough revenue to offset the infrastructure costs, known as the break-even point. Despite gaining a large number of users, it proved to be a challenge for the production and dissemination of hyperlocal advertisements to exceed this threshold. These factors that hinder revenue generation, combined with the high costs of the \name cloud network infrastructure, lead to challenges in supporting the \name ecosystem in developing regions.

% Additionally, based on the cloud network infrastructure and maintenance costs, we can estimate the amount of hyperlocal ad content required for the local ecosystem to be sustainable in developing regions. We assume that each content provider will produce an ad that, on average, targets and generates impressions from 5\% of the userbase. As the number of users increases, the percentage of ad content providers required to sustain the ecosystem decreases towards a more feasible value where the ecosystem becomes sustainable. If the amount of advertisement content is lower than the break-even point, then the CPM would need to be increased for the platform to be sustainable. Conversely, if the amount of ad content increases, then the CPM can be decreased below this threshold. 

% The Meta Ads Manager~\cite{metaads} uses historical data to calculate region-based estimates of the effectiveness of online advertisements for an application. The platform utilizes the target costs and target locations of the advertisement campaign as the inputs and produces estimates on the number of impressions that will be generated as a result of the campaign.  We observe that the expected ROAS for Meta Ads for the 2 regions is similar, however, a much larger number of impressions is required to achieve this property in developing regions.

}

{\bf \name Edge Deployment Challenges:} Deploying \name Edge servers in individual countries in partnership with local data center providers enabled us to significantly reduce the end-to-end latency in India and Kenya. However, managing edge deployments in emerging markets poses unique challenges on network and compute availability, power, maintenance, bandwidth and data cost constraints. In specific emerging markets like Bangladesh, we chose to co-locate our edge deployments at well-connected nodes in nearby countries. A major deployment challenge in Kenya for e.g. was that many users were using low end Android phones with an outdated Android (many users had Android 4.4 or less). The Kiberan community also faced several power outages, that led to periods where there was no \name usage. Another issue faced by the community was the cost of mobile access. All these users were on a \textit{Pay as You Go} and they were initially reluctant to use the \name platform as they feared usage of their mobile data (even though the platform offered a data saver mode). To tackle this issue the social enterprise provided free WiFi, for users to come over to their premise to use the platform. Although in terms of actual usage and interactions, the Kenyan deployment showcased that communities were extremely interested in using \name and there was an actual need, the issues surrounding mobile data usage combined with low end phones (with low memory) were barriers for large scale adoption. 

{\bf MAML: a Significant Bandwidth Saver:} Our lightweight representation of web content using MAML within the \name App allowed significant reductions in the pages' size and also eliminated the need for recursive content queries. 

{\bf Audio,Video in Local Languages:} As our deployments grew, we added support for content curation and consumption in several local languages since many users in our deployments were not English speaking users. Users preferred consumption in audio and video formats in local languages.

{\bf Local Content Summarization:} Locally relevant content on the Web is scattered across a broad array of web sources. As an extension, we built a content summarization service that was able to automatically crawl and summarize relevant local content on a specific topic from the Web and present a summarized view within each community. This feature enhanced user interactivity within specific communities that matched with local cultural interests.

{\bf Ad Monetization:} \name provided any user the ability to launch paid advertisements using the app. Unlike conventional monetization models using Google or Facebook Ads which provided limited payoffs, we promoted a bottom up advertising model where local businesses paid to advertise offers and deals within their own \name communities.

\textbf{Ethical Considerations: } This paper describes the product deployment experience for the \name platform. The findings presented in this paper are based on internal company data and the research presented in this paper has been authorized by the company's internal Institutional Review Board (IRB). Personal user data for the \name platform is protected under standard privacy laws. As such, we do not incorporate any individual's personal user data or present any user studies for this research.